**RESEARCH**            **Open Access**

# Fast restoration of natural images corrupted by high-density impulse noise

Hossein Hosseini[*] and Farokh Marvasti

**Abstract**

In this paper, we suggest a general model for the fixed-valued impulse noise and propose a two-stage method for high density noise suppression while preserving the image details. In the first stage, we apply an iterative impulse detector, exploiting the image entropy, to identify the corrupted pixels and then employ an Adaptive Iterative Mean filter to restore them. The filter is adaptive in terms of the number of iterations, which is different for each noisy pixel, according to the Euclidean distance from the nearest uncorrupted pixel. Experimental results show that the proposed filter is fast and outperforms the best existing techniques in both objective and subjective performance measures.

## Introduction

Images are often corrupted by impulse noise during acquisition and transmission; thus, an efficient noise suppression technique is required before subsequent image processing operations [1]. Median filter (MF) [2] is widely used in impulse noise removal methods due to its denoising capability and computational efficiency [3]. However, it is effective only for low noise densities. To overcome this drawback, many recent techniques [4-12] first detect the impulse locations and then filter the noisy pixels without processing the uncorrupted ones.

Various methods have been proposed to estimate the intensity values of noisy pixels. Some of the best existing methods are Opening-Closing Sequence (OCS) filter [4] based on mathematical morphology, Edge-Preserving Algorithm (EPA) [5] which adopts a directional correlation-dependent filtering technique, Switching-based Adaptive Weighted Mean filter (SAWM) [6] which replaces each noisy pixel with the weighted mean of its noise-free neighbors, Decision-based Average or Median filter (DAM) [7] which estimates the value of each noisy pixel using the average or median of the adjacent pixels, Cloud Model filter (CM) [8] which employs an uncertainty-based impulse detector and a weighted fuzzy mean filter and Fast and Efficient Median Filter (FEMF) [9] based on adaptive median filtering of noise-free pixels.

However, instead of two fixed values, impulse noise can be more realistically modeled by two fixed ranges that appear at both ends [10-12]. In [10], the Boundary Discriminative Noise Detection (BDND) is proposed to classify the pixels according to their intensities. The image restoration is done by using the switching median filter. A Noise Fading Technique (NFT), presented in [11], first applies an impulse detector and then employs a pixel restoring median filter for denoising the corrupted pixels in an iterative manner. In [12], a Two-State Switching Median filter (TSSM) is proposed which uses a noise intensity identification scheme to detect the corrupted pixels. The filtering scheme is chosen from two different switching median filters.

In this paper, we suggest a general model for impulse noise which assumes that impulses can take any subset of the entire grey-values dynamic range. Noisy pixels are detected based on iterative measuring the image entropy and restored using an Adaptive Iterative Mean filter (AIM). The proposed filter alters the corrupted pixels in a different number of iterations according to their Euclidean distance from the nearest uncorrupted pixel.

The rest of this paper is organized as follows. Section II reviews the impulse noise models and describes the suggested one. Our algorithm is presented in Section III. The experimental results and comparisons are provided in Section IV and Section V concludes the paper.

* Correspondence: h_hosseini@alum.sharif.edu
Department of Electrical Engineering, Sharif University of Technology,
Advanced Communication Research Institute (ACRI), Tehran, Iran

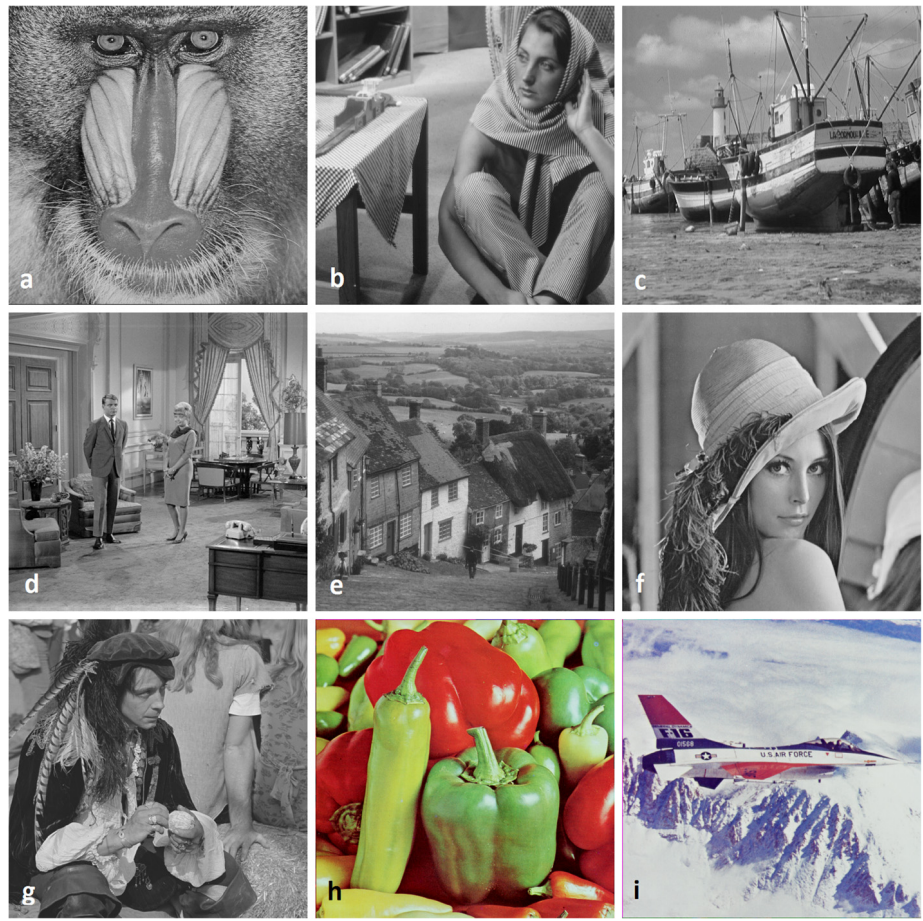

**Figure 1 Test images.** (**a**) *Baboon*, (**b**) *Barbara*, (**c**) *Boat*, (**d**) *Couple*, (**e**) *Hill*, (**f**) *Lena*, (**g**) *Man*, (**h**) *Peppers*, (**i**) *F16*.

### Impulse noise model

Two common types of the impulse noise are the Fixed-Valued Impulse Noise (FVIN), also known as Salt-and-Pepper Noise (SPN), and the Random-Valued Impulse Noise (RVIN). They differ in the possible values which noisy pixels can take. The FVIN is commonly modeled by

$$y_{i,j} = \begin{cases} \{0, 255\} & \text{with probabbility} \quad p \\ x_{i,j} & \text{with probabbility} \quad 1-p \end{cases} \quad (1)$$

where $x_{i,j}$ and $y_{i,j}$ denote the intensity value of the original and corrupted images at coordinate (*i,j*), respectively and *p* is the noise density. This model implies that the pixels are randomly corrupted by two fixed extreme values, 0 and 255 (for 8-bit grey-scale images), with the same probability.

In [10], a model is considered as below:

$$y_{i,j} = \begin{cases} [0, m) & \text{with probability} \quad p_1 \\ x_{i,j} & \text{with probability} \quad 1-p \\ (255-m, 255] & \text{with probability} \quad p_2 \end{cases} \quad (2)$$

where $p = p_1 + p_2$. We refer to this model as Fixed-Range Impulse Noise (FRIN). In this model, instead of two fixed *values*, two fixed *ranges* at both ends with the length of *m* are assumed to be impulse noise values. Also the densities of low-intensity and high-intensity impulse noise might be unequal.

**Table 1 The entropy of test images**

| Image | Entropy |
|---|---|
| Baboon | 7.36 |
| Barbara | 7.63 |
| Boat | 7.19 |
| Couple | 7.06 |
| Hill | 7.48 |
| Lena | 7.45 |
| Man | 7.19 |
| F16 | 6.58 |
| Peppers | 7.30 |
| **Average** | 7.25 |

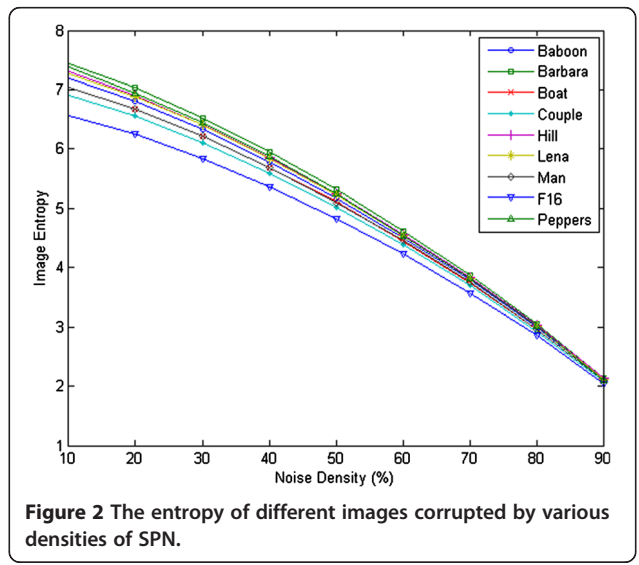

**Figure 2** The entropy of different images corrupted by various densities of SPN.

In real applications, impulse values can be any subset of the entire range of grey-values. Thus, we introduce a more general model, as:

$$y_{i,j} = \begin{cases} S & \text{with probability} \quad p \\ x_{i,j} & \text{with probability} \quad 1-p \end{cases} \quad (3)$$

where $S$ is the set of impulse noise values with $k$ elements chosen from the set [0 255]. The impulse probabilities don't have to be equal. We refer to this model as General Fixed-Valued Impulse Noise (GFN) or Multi-Valued Impulse Noise (MVIN). Models described in Equations (1) and (2) are also included in the GFN model by choosing the appropriate set $S$ as $S_1 = \{0,255\}$ and $S_2 = \{0,1,\ldots,m-1,255-m+1,255-m+2,\ldots,255\}$, respectively. The GFN can bridge the FVIN and the RVIN; if we choose $s = [0\ 255]$ and all values have the same probability, the RVIN will be obtained.

## The proposed method

Many recent image denoising techniques consist of two stages: first detecting the corrupted pixels and then estimating their original values. Traditional impulse detectors fail, when the impulses have few values, distributed in the entire range of grey-values. For the GFN model, we need an Impulse Value Detector (IVD) to determine the noise values. The method is described below:

*A-Impulse Value Detector (IVD):*
The entropy of the image $y$ is defined as follows:

$$entropy(y) = -\sum_{i=0}^{255} p_i \log p_i \quad (4)$$

where $p_i$ is the probability of the grey-value $i$ and can be interpreted as the normalized histogram of the image. In natural images, unlike the Gaussian noise, the impulse noise significantly decreases the image entropy.

The impulse value detector, iteratively, detects and removes the impulse grey-values. A grey-value is detected as an impulse if the corresponding pixels have the lowest correlation with their neighbors. After each iteration, by removing an impulse value, the image entropy increases. The process continues until the entropy becomes larger than the entropy threshold, thus it ensures that there are no more impulse values in the image.

The details of the Impulse Value Detector (IVD) are described below:

1. Construct the Mask matrix with the same size as the image. The Mask is zero in the corresponding image pixels, which are determined as corrupted and is one otherwise.

2. Construct the detailed part of the image as follows:

$$y^D = \left| \frac{y * h}{\text{Mask} * h} - y \right| \quad (5)$$

where the symbols (*) and |.| denote the convolution and the absolute value function, respectively and $h$ is the filtering window defined as:

$$h = \begin{bmatrix} 0 & 1 & 0 \\ 1 & 0 & 1 \\ 0 & 1 & 0 \end{bmatrix} \quad (6)$$

3. The noisy part of the image is determined as below:

$$y^N{}_{i,j} = \begin{cases} 0 & \text{if } y^D{}_{i,j} \leq \text{correlation threshold} \\ y_{i,j} & \text{otherwise} \end{cases} \quad (7)$$

**Table 2 Restoration results in PSNR (dB) for image *Lena* corrupted by various densities of SPN**

|  | 50% | 60% | 70% | 80% | 90% |
|---|---|---|---|---|---|
| **Median Filter** | 15.27 | 12.35 | 10.01 | 8.14 | 6.64 |
| **OCS** [4] | 30.63 | 30.55 | 29.71 | 27.95 | 25.58 |
| **EPA** [5] | 34.10 | 32.66 | 31.03 | 29.01 | 26.04 |
| **SAWM** [6] | 33.82 | 32.32 | 30.69 | 28.84 | 26.17 |
| **DAM** [7] | 32.78 | 31.24 | 29.68 | 27.95 | 25.54 |
| **CM** [8] | 33.85 | 32.11 | 30.71 | 28.59 | 26.02 |
| **FEMF** [9] | 33.28 | 31.64 | 30.18 | 28.47 | 25.94 |
| **AIM** [Proposed] | **34.17** | **32.82** | **31.40** | **29.82** | **27.52** |

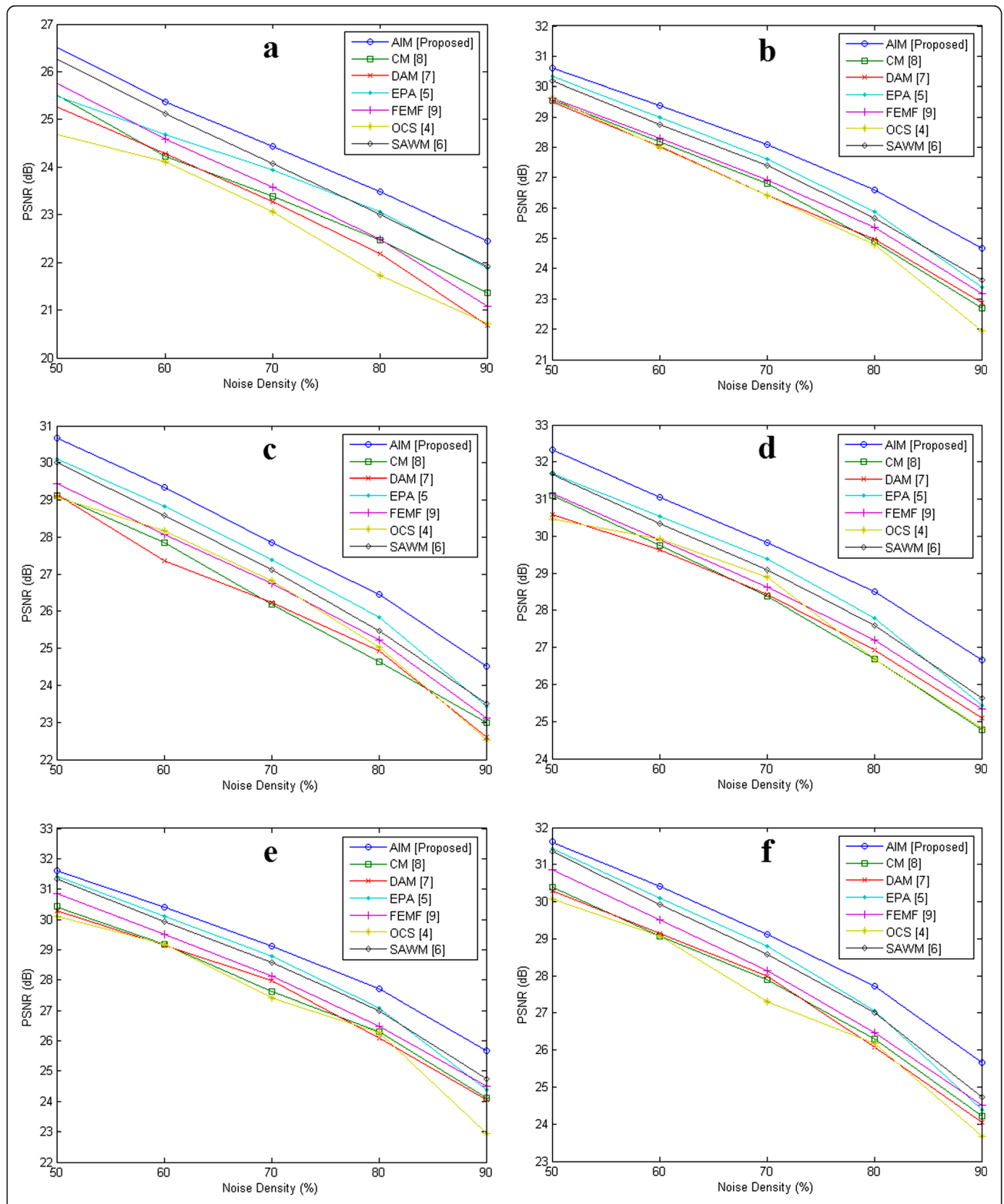

Figure 3 **Comparison of restoration results in PSNR (dB) for different images corrupted by various densities of SPN.** (**a**) *Barbara*, (**b**) *Boat*, (**c**) *Couple*, (**d**) *Hill*, (**e**) *Man*, (**f**) *F16*.

**Table 3 Restoration results in PSNR (dB) for different images corrupted by 80% SPN**

|  | Median filter | OCS [4] | EPA [5] | SAWM [6] | DAM [7] | CM [8] | FEMF [9] | AIM [Proposed] |
|---|---|---|---|---|---|---|---|---|
| *Baboon* | 8.15 | 20.48 | 21.30 | 20.95 | 20.34 | 20.53 | 20.46 | **21.44** |
| *Barbara* | 7.90 | 21.69 | 23.08 | 22.96 | 22.09 | 22.80 | 22.44 | **23.49** |
| *Boat* | 8.12 | 24.73 | 25.87 | 25.68 | 25.01 | 25.63 | 25.48 | **26.57** |
| *Couple* | 8.19 | 24.87 | 25.72 | 25.52 | 25.06 | 25.51 | 25.24 | **26.45** |
| *Hill* | 8.07 | 26.63 | 27.83 | 27.62 | 27.12 | 27.33 | 27.27 | **28.51** |
| *Lena* | 8.14 | 27.95 | 29.01 | 28.84 | 27.95 | 28.59 | 28.47 | **29.82** |
| *Man* | 8.08 | 26.06 | 27.11 | 26.85 | 26.24 | 26.39 | 26.52 | **27.70** |
| *F16* | 7.57 | 24.91 | 26.02 | 26.30 | 25.72 | 25.96 | 25.98 | **27.09** |
| *Peppers* | 7.64 | 24.17 | 25.63 | 25.56 | 25.17 | 25.15 | 25.40 | **26.58** |
| **Average** | 7.98 | 24.61 | 25.73 | 25.59 | 24.97 | 25.32 | 25.25 | **26.41** |

Calculate the histogram of the positive elements of the image $y^N$ and declare the most frequent grey-value $g_{max}$ as an impulse value.

4. In the image $y$, replace all the pixels with values equal to $g_{max}$ with zero.

5. If the entropy of positive elements of the image $y$ is larger than the entropy threshold, stop the iterations, otherwise go to step 1.

B. Image Restoration Using the AIM Filter:

Unlike many techniques which increase the size of the window, until all corrupted pixels are detected and restored, our approach is to employ a small filter, iteratively. This approach has two advantageous: 1) better results, because of stronger correlation among adjacent pixels and 2) less complexity, because of implementation using linear shift-invariant filters.

To restore noisy pixels, we first compute the Distance Transform (DT) of the image. Each element of DT contains the Euclidean distance of the corresponding image pixel with the nearest uncorrupted pixel. For Euclidean distance calculation, we employed the fast algorithm described in [13].

The image $y^0$ is the input for the AIM filter. In this image, each noisy pixel takes the grey-value of its nearest uncorrupted pixel. In the AIM filter, the noisy pixels which their corresponding elements of DT have larger value, will be modified in more iterations. Also, Mask$^0$ is the all one matrix with the same size as the image and $p$

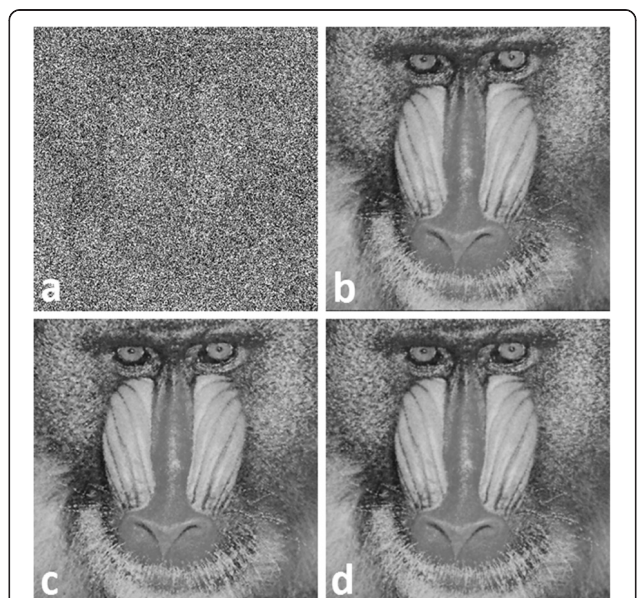

**Figure 4 Reconstructed images using different filters for image *Baboon* corrupted by 80% SPN.** (**a**) Corrupted image (6.54 dB), (**b**) EPA [7] (21.33 dB), (**c**) SAWM [8] (20.92 dB) and (**d**) AIM [Proposed] (21.58 dB).

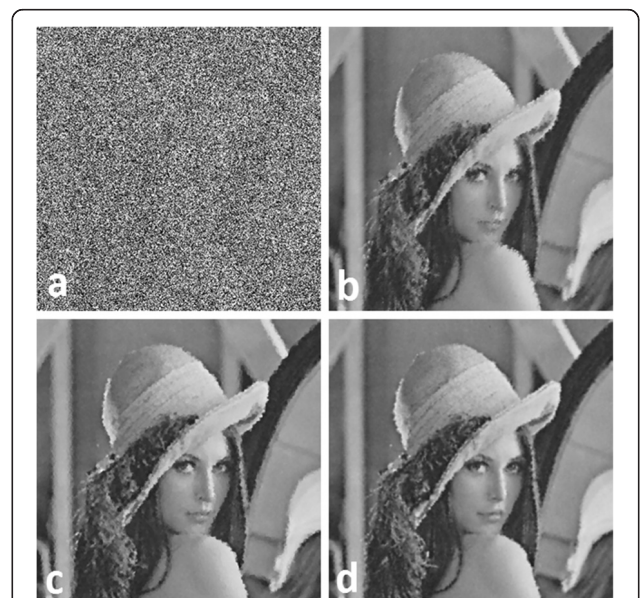

**Figure 5 Reconstructed images using different filters for image *Lena* corrupted by 90% SPN.** (**a**) Corrupted image (5.91 dB), (**b**) EPA [7] (26.03 dB), (**c**) SAWM [8] (26.15 dB) and (**d**) AIM [Proposed] (27.45 dB).

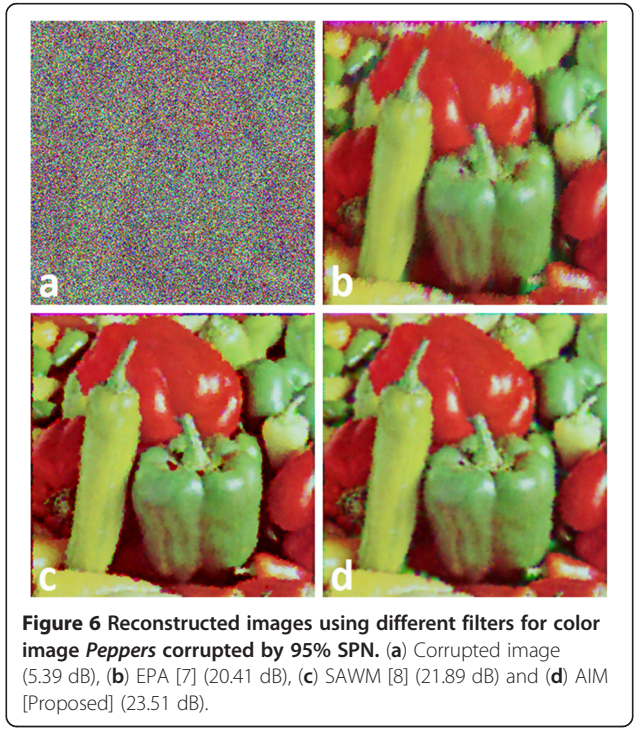

**Figure 6 Reconstructed images using different filters for color image *Peppers* corrupted by 95% SPN.** (**a**) Corrupted image (5.39 dB), (**b**) EPA [7] (20.41 dB), (**c**) SAWM [8] (21.89 dB) and (**d**) AIM [Proposed] (23.51 dB).

is the probability of corrupted pixels, which are determined by IVD.

The fast implementation of the AIM filter is described below:

1. Iteratively update the image and the Mask matrix. Start from $k=1$ and continue until $\frac{k}{3p}$ is greater than the maximum element of DT.

$$y^k_{i,j} = \begin{cases} y^{k-1}_{i,j} & \text{if } DT_{i,j} < \frac{k}{3p} \\ y^{k-1}_{i,j} * h & \text{otherwise} \end{cases} \quad (8)$$

$$Mask^k_{i,j} = \begin{cases} Mask^{k-1}_{i,j} & \text{if } DT_{i,j} < \frac{k}{3p} \\ Mask^{k-1}_{i,j} * h & \text{otherwise} \end{cases} \quad (9)$$

2. Restored image is obtained by pixel-wise division of the final image and Mask matrix as below:

$$\text{Restored Image} = \frac{y^{Final}}{Mask^{Final}} \quad (10)$$

Some uncorrupted pixels which have one of impulse values are identified as corrupted by IVD. They are likely to be reproduced in the restoration stage. However, to avoid false alarms, at the end and for each pixel, if the difference between the estimated and original pixel values is lower than the correlation threshold, we replace it with the original value.

**Table 4 Runtime in seconds for various densities of SPN in the MATLAB environment**

|  | 50% | 60% | 70% | 80% | 90% |
|---|---|---|---|---|---|
| **Median Filter** | 0.006 | 0.006 | 0.006 | 0.006 | 0.006 |
| **OCS** [4] | 13.1 | 13.5 | 13.8 | 14.0 | 14.4 |
| **EPA** [5] | 0.64 | 0.74 | 0.82 | 0.86 | 0.88 |
| **SAWM** [6] | 7.0 | 7.7 | 8.8 | 10.1 | 10.9 |
| **DAM** [7] | 2.1 | 2.6 | 2.9 | 3.4 | 4.1 |
| **CM** [8] | 10.8 | 16.9 | 18.4 | 23.0 | 28.5 |
| **FEMF** [9] | 2.3 | 2.9 | 3.2 | 3.7 | 4.3 |
| **AIM** [Proposed] | **0.17** | **0.19** | **0.22** | **0.29** | **0.38** |

### Simulation results

The proposed method is compared with the best existing techniques for SPN and FRIN removal. Comparisons include the quantitative evaluation, the visual quality and the time complexity. Simulations are carried out on $512 \times 512$ test images shown in Figure 1. In simulations, a color image is considered as three separate grey-scale images. The Peak Signal-to-Noise Ratio (PSNR) is employed for objective performance measurement. To make a reliable comparison, each method is run 20 times with different impulse noise patterns and the result is obtained by averaging over all experiments.

Table 1 lists the entropy of test images and Figure 2 shows the entropy of different images corrupted by SPN with various noise densities. Natural 8-bit grey-scale images have the entropy around 7. In simulations, the entropy threshold is set to 6 to avoid any miss detection and the correlation threshold is set to 8. We consider two cases: SPN and FRIN. The results are provided below:

A. *Salt & Pepper Noise (SPN):*

Table 2 lists the restoration results of different techniques for image *Lena* corrupted by various densities of SPN. Figure 3 evaluates the results for different images. In Table 3, the comparisons are done for different images corrupted by 80% SPN. The results reveal that the AIM filter performs considerably better in high noise densities. Figures 4, 5 and 6 demonstrate reconstructed images *Baboon*, *Lena* and *Peppers* corrupted by 80%, 90% and 95% SPN, respectively. Apparently, the AIM

**Table 5 Restoration results in PSNR (dB) for image *Lena* corrupted by 80% FRIN**

| Noise range | | PSNR | | | |
|---|---|---|---|---|---|
| Low-intensity | High-intensity | BDND [10] | NFT [11] | TSSM [12] | AIM [Proposed] |
| [0 9] | [246 255] | 18.79 | 25.16 | 28.55 | **29.85** |
| [0 29] | [226 255] | 19.07 | 23.67 | 28.42 | **29.78** |
| [0 49] | [206 255] | 17.31 | 22.83 | 25.16 | **26.22** |

filter outperforms other techniques in preserving the image details.

Also, Table 4 provides the running time of different filters in the MATLAB environment. For these filters, the computational complexity depends on the noise density, rather than the image. The time complexity can be measured with respect to the standard median filter implemented in the MATLAB software. The results show that the proposed filter is very efficient for high density impulse noise removal.

B. Fixed-Range Impulse Noise (FRIN):

Traditional methods for SPN removal fail to restore images corrupted by FRIN. Thus, in this case we compare the proposed filter only with appropriate techniques, i.e. BDND, NFT and TSSM.

Table 5 compares the restoration results for image Lena corrupted by 80% FRIN with different low-intensity and high-intensity impulse ranges. It can be seen that the proposed method significantly outperforms other techniques. The AIM filter can remove FRIN, even if the low and high intensities have different probabilities.

## Conclusion

In this paper, we introduced a more realistic model for the impulse noise called General Fixed-Valued Impulse Noise (GFN). This model implies that, instead of fixed values or ranges, impulses can take any subset of the entire grey-values dynamic range. This model requires a more complex impulse detector, because the noise values do not necessarily locate at the low and high-intensities. We proposed a procedure for impulse value detection using the image entropy. For image restoration, an Adaptive Iterative Mean (AIM) filter is presented. In this filter, the noisy pixels which are farther than their nearest uncorrupted pixel, will be modified in more iterations. The AIM filter outperforms the best existing techniques for SPN and FRIN removal. The proposed method is fast and quite suitable for real-time applications.